\documentclass[runningheads]{llncs}

\newcommand{\x}{\mathbf{x}}
\newcommand{\y}{\mathbf{y}}
\usepackage{graphicx}
\usepackage{xcolor}
\usepackage{amsmath,enumitem,mathtools}
\usepackage{amssymb}
\usepackage{hyperref}
\usepackage{cite,cleveref}
\begin{document}
%

\title{A Machine-Learning-Based Importance Sampling Method to Compute Rare Event Probabilities\thanks{This material was based upon work
	supported by the U.S. Department of Energy, Office of Science,
	Office of Advanced Scientific Computing Research (ASCR) under
	Contract DE-AC02-06CH11347.}}

%
%
\author{Vishwas Rao\inst{1} \and
Romit Maulik\inst{1} \and
Emil Constantinescu\inst{1} \and Mihai Anitescu \inst{1,2}}
\authorrunning{V. Rao et al.}
%
\institute{Argonne National Laboratory, Lemont, IL, USA \and
University of Chicago, Chicago, IL, USA\\
\email{\{vhebbur,rmaulik,emconsta,anitescu\}@anl.gov}}
\maketitle              
\begin{abstract}
    We develop a novel computational method for evaluating the extreme excursion probabilities arising from random initialization of nonlinear  dynamical  systems. The  method  uses  excursion probability theory to formulate a sequence of Bayesian inverse problems that, when solved, yields the biasing distribution. Solving multiple Bayesian inverse problems can be expensive; more so in higher dimensions. To alleviate the computational cost, we build machine-learning-based  surrogates  to  solve the Bayesian inverse problems that give rise to the biasing distribution. This biasing distribution can then be used in an importance sampling procedure to estimate the extreme excursion probabilities.  

\end{abstract}
\section{Motivation}\label{sec:intro}
Characterizing high-impact rare and extreme events such as hurricanes, tornadoes, and cascading power failures are of great social and economic importance. Many of these natural phenomena and engineering systems can be modeled by using dynamical systems. The models representing these complex phenomena are approximate and have many sources of uncertainties. For example, the exact initial and boundary conditons or the external forcings that are necessary to fully define the underlying model might be unknown. Other parameters that are set based on experimental data may also be uncertain or only partially known. A probabilistic framework is generally used to formulate the problem of quantifying various uncertainties in these complex systems. By definition, the outcomes of interest that correspond to high-impact rare and extreme events reside in the tails of the probability distribution of the associated event space. Fully characterizing the tails  requires resolving high-dimensional integrals over irregular domains. The most commonly used method to determine the probability of rare and extreme events is  Monte Carlo simulation (MCS). Computing rare-event probabilities via MCS involves generating several samples of the random variable and calculating the fraction of the samples that produce the outcome of interest. For small probabilities, however, this process is expensive. For example, consider an event whose probability is around $10^{-3}$ and  the underlying numerical model for the calculation  requires ten minutes per simulation. With MCS, estimating the probability of such an event to an accuracy of $10\%$  will require two years of serial computation. Hence,  alternative methods are needed  that are computationally efficient.

Important examples of extreme events are rogue waves in the ocean  \cite{Dysthe_2008}, hurricanes, tornadoes \cite{Ross_2003}, and power outages \cite{Atputharajah_2009}. The motivation for this work comes from the rising concern surrounding transient security in the presence of uncertain initial conditions  identified by North American Electric Reliability Corporation in connection with its long-term reliability assessment \cite{nerc2017ltra}. The problem can be mathematically formulated as a dynamical system with uncertain initial conditions. In this paper, the aim is to compute the extreme excursion probability: the probability that the transient due to a sudden malfunction exceeds preset safety limits. Typically, the target safe limit exceedance probabilities are in the range $10^{-4}--10^{-5}$.
We note that the same formulation is applicable in other applications such as data assimilation, which is used extensively for medium- to long-term weather forecasting. For example, one can potentially use the formulation in this paper to determine the likelihood of temperature levels at a location exceeding certain thresholds or the likelihood of precipitation levels exceeding safe levels in a certain area.

In \cite{Rao_2020}, we presented an algorithm that uses ideas from excursion probability theory to evaluate the probability of extreme events \cite{adler2010geometry}. In particular we used Rice's formula \cite{Rice_1944}, which  was developed to estimate the average number of upcrossings for a generic stochastic process. Rice's formula is given by
\begin{equation}\label{eqn:RicesFormula}
        \mathbb{E} \left\{N^{+}_u(0,T) \right \} = \displaystyle \int_{0}^{T} \, \int_{0}^{\infty}\, y \varphi_t(u,y)\, \mathrm{d}y\, \mathrm{d}t\, ,
\end{equation}
where the left-hand side denotes the expected number of upcrossings of level u, y is the derivative of the stochastic process in a mean squared sense, and $\varphi_t(u,y)$ represents the joint probability distribution of the process and its derivative.
In this paper, we build on our recent algorithm \cite{Rao_2020}, 
which we used to construct an importance biasing distribution (IBD) to accelerate the computation of extreme event probabilities. A key step in the algorithm presented in  \cite{Rao_2020} involves solving multiple Bayesian inverse problems, which can be expensive in high dimensions. Here, we propose to use machine-learning-based surrogates to obtain the inverse maps and hence alleviate the computational costs.

\subsection{Mathematical setup and overview of the method}
The mathematical setup used in this paper consists of a nonlinear dynamical system that is excited by a Gaussian initial state and that results in a non-Gaussian stochastic process. We are interested in estimating the probability of the stochastic process exceeding a preset threshold. Moreover, we wish to estimate the probabilities when the underlying event of the process exceeding the threshold is a \textit{rare event}. The rare events typically lie in the tails of the underlying event distribution. To characterize the tail of the resulting stochastic process, we use ideas from theory excursion probabilities \cite{adler2010geometry}. Specifically, we use Rice's formula \eqref{eqn:RicesFormula} to estimate the expected number of upcrossings of a stochastic process. For a description of the mathematically rigorous settings used for the rare event problem, we refer  interested readers to \cite[\S 2]{Rao_2020} and references therein. 

Evaluating $\varphi_t(u,y)$, the joint probability distribution of the stochastic process and its derivative, is central to evaluating the integral in Rice's formula. However, $\varphi_t(u,y)$ is analytically computable  only for Gaussian processes. Since our setup results in a non-Gaussian stochastic process, we linearize the nonlinear dynamical system variation around the trajectories starting at the mean of the initial state. We  thus obtain a Gaussian approximation to the system trajectory distribution.
In \cite{Rao_2020}, we solve a sequence of Bayesian inverse problems to determine a biasing distribution to accelerate the convergence of the probability estimates. Forr high-dimensional problems, however, solving multiple Bayesian inverse problems can be expensive. In this work, we propose to replace multiple solutions to Bayesian inverse problems with machine-learning-based surrogates to alleviate the computational burden.

\subsection{Organization}The rest of the paper is organized as follows. In \S \ref{sec:existing_literature} we review the existing literature for estimating rare event probabilities. In \S\ref{sec:mcmc-based-sampling} we reformulate the problem of determining the IBD as a Bayesian inference problem, and in \S\ref{sec:ml_inverse_maps} we develop a machine-learning-based surrogate to approximate the solution to the Bayesian inference problem. In \S\ref{sec:numerical_experiments} we demonstrate this  methodology  on a simple nonlinear dynamical system excited by a Gaussian distribution. In \S\ref{sec:conc} we present our conclusions and potential future  research directions.

\section{Existing literature}\label{sec:existing_literature}
\subsection{Monte Carlo and importance sampling}Most of the existing methods to compute the probabilities of rare events use MCS directly or indirectly. The MCS approach was developed by Metropolis and his collaborators to solve problems in mathematical physics \cite{Metropolis_1949}. Since then, it has been used in a variety of applications \cite{Liu_2008B, Robert_2005B}. When evaluating rare event probabilities, the MCS method basically counts the fraction of the random samples that cause the rare event. For a small probabilty $P$ of the underlying event, the number of samples required to obtain an accuracy of $\epsilon \ll 1$ is $\mathcal{O}(\epsilon^{-2} P^{-1})$. Hence MCS becomes impractical for estimating rare event probabilities. 

A popular sampling technique that is employed to compute rare event probabilties is importance sampling (IS). IS is a variance reduction technique  developed in the 1950s \cite{Kahn_1953} to estimate the quantity of interest by constructing estimators that have smaller variance than MCS. In MCS, simulations from most of the samples do not result in the rare event and hence do not play a part in probability calculations.  IS, instead, uses problem-specific information to construct an IBD; computing the rare event probability using the IBD requires fewer samples. Based on this idea, several techniques for constructing IBDs have been developed \cite{Bucklew_2013B}. For a more detailed treatment of IS, we direct  interested readers to \cite{Dunn_2011B, Asmussen_2007B}.
One of the major challenges involved with importance sampling is the construction of an IBD that results in a low-variance estimator. We note that the approach may sometimes be inefficient for high-dimensional problems \cite{Katafygiotis_2008}. A more detailed description of MCS and IS in the context of rare events can be found in \cite[\S 2]{Rao_2020} and references therein.
\subsection{Nested subset methods}\label{sec:nested} Other methods use the notion of conditional probability over a sequence of nested subsets of the probability space of interest. For example, one can start with the entire probability space and progressively shrink to the region that corresponds to the rare event. Furthermore, one can use the notion of conditional probability to factorize the event of interest as a product of conditional events. Subset simulation (SS) \cite{Au_2001} and splitting methods \cite{Kahn_1951} are ideas that use this idea. Several modifications and improvements have been proposed to both SS \cite{Ching_2005, Ching_2005_2, Katafygiotis_2005, Zuev_2012, Bect_2017} and splitting methods \cite{Botev_2012, Beck_2016}. Evaluating the conditional probabilities forms a major portion of the computational load.  Compute the conditional probabilities for different nested subsets concurrently is nontrivial. 
\subsection{Other approaches} Large deviation theory (LDT) is an efficient approach for estimating rare events in cases when the event space of interest is dominated by few elements such as  rogue waves of a certain height. LDT also has been used to estimate the probabilities of extreme events in dynamical systems with random components \cite{Dematteis_2018, Dematteis_2019}. A sequential sampling strategy has been used to compute extreme event statistics \cite{Mohamad_2018, Mohamad_2018B}. 

\section{A Bayesian inference formulation to construct IBD}\label{sec:mcmc-based-sampling}
Most of the work in this section is a review of our approach described in \cite[\S 3]{Rao_2020}. Here we reformulate the problem of constructing an IBD as a sequence of Bayesian inverse problems.
Consider the following dynamical system,
\begin{align}\label{eqn:dynamics}
        \x' &=  f(t,\x)\,, \quad t  =[0, T] \\
        \x(0) &= \x_0\,, \quad \x_0 \sim p\,, \quad \x \in \Omega\,, \nonumber
\end{align}
%
where $\mathbf{c}$ is a canonical basis vector and $\x_0$, the initial state of the system, is uncertain and has a probability distribution $p$. The problem of interest is to estimate the probability that $ \mathbf{c}^{\top} \mathbf{x}(t)$ exceeds the level $u$ for $t\in \lbrack 0, T \rbrack$. That is, we seek to estimate the following {\textit{excursion probability}},
\begin{align}\label{eqn:problem_of_interest}
P_T(u) \coloneqq \mathbb{P}\left( \sup_{0 \leq t \leq T} \mathbf{c}^{\top}\mathbf{x}(t, \x_0) \geq u\,, ~~t\in \lbrack 0, T \rbrack \right)\,,
\end{align}
where $\x(t, \x_0)$ represents the solution of the dynamical system \eqref{eqn:dynamics} for a given initial condition $\x_0$.
We note that 
\begin{align}\label{eqn:probability_conditional}
   P_{T}(u) = \mu(\Omega(u))\,,
\end{align}
where $\mu$ is the respective measure transformation subject to \eqref{eqn:dynamics} and $\Omega(u) \subset \Omega$ represents the \textit{excursion set}
\begin{align}
\Omega(u) \coloneqq \left \{\x_0 : \sup_{0 \leq t \leq T}\mathbf{c}^{\top}\x(t,\x_0) \geq u  \right \}\,.
\end{align}
Hence, estimating $\Omega(u)$ will help us in estimating the excursion probability $P_T(u)$. In general, however,  estimating the excursion set $\Omega(u)$ analytically is difficult. Rice's formula, \eqref{eqn:RicesFormula} gives us insights about the excursion set and can be used to construct an approximation to the excursion set.

Recall that in Rice's formula (1), 
$\varphi_t(u,y)\,$ represents the joint probability density of $\mathbf{c}^{\top}\x$ and its derivative $\mathbf{c}^{\top}\x'$ for an excursion level $u$. The right-hand side of (1) can be interpreted as the summation of all times and slopes at which  an excursion occurs. One can sample from $y \varphi_t(u,y)$ to obtain a slope-time pair $(y_i, t_i)$ at which the sample paths of the stochastic process cause an excursion. Now consider the map $\mathcal{G}: \mathbb{R}^{d\times 1} \rightarrow \mathbb{R}^2$ that evaluates the vector $\displaystyle \begin{bmatrix}\mathbf{c}^{\top}\x(t) \\ \mathbf{c}^{\top}\x'(t) \end{bmatrix}$ based on the dynamical system \eqref{eqn:dynamics}, given an initial state $\x_0$ and a time $t$. By  definition of the excursion set $\Omega(u)$, there exists an element $\x_i \in \Omega(u)$ that satisfies the following relationship,
\begin{align}\label{eqn:forwardmapG}
        \mathcal{G}(\x_i, t_i) = \displaystyle \begin{bmatrix} u + \varepsilon_i\\ y_i \end{bmatrix} \,,   
        \end{align}
where $\varepsilon > 0$. We can use this insight to construct an approximation of $\Omega(u)$ by constructing the preimages of multiple slope-time pairs. Observe that the problem of finding the preimage of a sample $(y_i,t_i)$ is ill-posed since there could be multiple $\x_i$'s that map to $\begin{bmatrix} u + \varepsilon_i\\ y_i \end{bmatrix}$ at $t_i$ via operator $\mathcal{G}$. We define the set 
\begin{align}\label{eqn:preimages}
        X_i \coloneqq \left \{\x_i \in \Omega : \mathcal{G}(\x_i, t_i) = \displaystyle \begin{bmatrix} u + \varepsilon_i \\ y_i \end{bmatrix}  \right \}\,,
\end{align}
and an approximation $\widehat{\Omega}(u)$ to $\Omega(u)$ can be written as
\begin{align}\label{eqn:approximate_omega_u}
        \widehat{\Omega}(u) \coloneqq \bigcup_{i=1}^N X_i\,.
\end{align}
Note that the approximation \eqref{eqn:approximate_omega_u} improves as we increase $N$. For a discussion on the choice of $\varepsilon_i$, we refer interested readers to \cite[\S 3.3]{Rao_2020}.

The underlying computational framework to approximate $\widehat{\Omega}(u)$ consists of the following stages:
\begin{itemize}
        \item Draw samples from unnormalized $y \varphi_t(u,y)\,$
        \item Find the preimages of these samples to approximate $\Omega(u)$.
\end{itemize}

We use MCMC to draw samples from unnormalized $y \varphi_t(u,y)\,$. We note that irrespective of the size of the dynamical system, $y \varphi_t(u,y)\,$ represents an unnormalized density in two dimensions; hence, using MCMC is an effective means , draw samples from it. Drawing samples from $y \varphi_t(u,y)\,$ requires evaluating it repeatedly, and in the following section we discuss the means to do so. 

\subsection{Evaluating $y \varphi_t(u,y)\,$}\label{sec:evaluatevarphi} 
We note that $y \varphi_t(u,y)\,$ can be evaluated analytically only for special cases. Specifically, when $\varphi_t(u,y)$ is a Gaussian process, then the joint density function $y \varphi_t(u,y)\,$ is analytically computable. Consider the dynamical system described by \eqref{eqn:dynamics}. When $p$ is Gaussian and $f$ is linear, we have
\begin{align}\label{eqn:linearGaussian}
        \x' = A\, \x(t) + b\,, \quad \x(t_0) = \x_0\,, \quad \x_0 \sim \mathcal{N}(\overline{\x}_0, \Sigma)\,.
\end{align}
Assuming $A$ is invertible, $\x(t)$ can be written as
\begin{align}\label{eqn:closedform}
\x(t) = \exp(A(t-t_0)) \, \x_0  - \left(I  - \exp(A(t-t_0))\right)A^{-1} b\,,
\end{align}
where $I$ represents an identity matrix of the appropriate size. Given that $\x_0$ is normally distributed, it follows that $\x(t)$ is a Gaussian process:
\begin{align}\label{eqn:gpx}
        &\x(t) \sim \mathcal{GP} \left(\overline{\x}, {\rm cov}_{\x} \right ) \,, \text{ where }\\
     &\nonumber     \overline{\x} = \exp(A(t-t_0)) \overline{\x}_0 - \left(I  - \exp(A(t-t_0))\right)A^{-1} b \, \text{ and } \\
     &\nonumber   {\rm cov}_{\x} = \exp(A(t-t_0)) \Sigma \left(\exp(A(t-t_0))\right)^\top\,.
\end{align}
The joint probability density function (PDF)
 of $\mathbf{c}^{\top}\mathbf{x}(t)$ and $\mathbf{c}^{\top}\mathbf{x}'(t)$ is given by \cite[equation 9.1]{3569}
\begin{align}\label{eqn:gpvarphi}
        \begin{bmatrix}
                \mathbf{c}^{\top}\x \\
                \mathbf{c}^{\top}\x'
        \end{bmatrix}
        \sim \mathcal{GP}\left(\overline{\x}^{\varphi}, \begin{bmatrix}
                \mathbf{c}^{\top}\Phi\mathbf{c} & \mathbf{c}^{\top}\Phi A^{\top}\mathbf{c} \\
                                            \mathbf{c}^{\top}A \Phi^{\top}\mathbf{c}& \mathbf{c}^{\top}A\Phi A^{\top}\mathbf{c}
                                          \end{bmatrix}
        \right)\,,  
\end{align}
where $$\overline{\x}^{\varphi} \coloneqq \begin{bmatrix} \mathbf{c}^{\top} \overline{\x} \\ \mathbf{c}^{\top} (A\overline{\x} + b)\end{bmatrix}$$ and $$\Phi \coloneqq \exp(A(t-t_0)) \Sigma \left(\exp(A(t-t_0))\right)^\top\,.$$
We now can evaluate $y \varphi_t(u,y)\,$ for arbitrary values of $u_i$, $y_i$, and $t_i$ as 
\begin{align}\label{eqn:compute_varphi}
y_i \varphi_{t_i}(u_i,y_i) = \frac{y_i}{2 \pi \mid \Upsilon \mid}\exp\left(-\frac{1}{2} \left \| \begin{bmatrix} u_i \\ y_i \end{bmatrix} - \overline{\x}^{\varphi}\right \|^2_{\Upsilon^{-1}}\right)\,,
\end{align}
where $\Upsilon \coloneqq  \begin{bmatrix}
        \mathbf{c}^{\top}\Phi\mathbf{c} & \mathbf{c}^{\top}\Phi A^{\top}\mathbf{c} \\
                                    \mathbf{c}^{\top}A \Phi^{\top}\mathbf{c}& \mathbf{c}^{\top}A\Phi A^{\top}\mathbf{c}
                                  \end{bmatrix}$ and $\mid \Upsilon \mid$ denotes the determinant of $\Upsilon$. Note that the right-hand side in \eqref{eqn:compute_varphi} is dependent on $t_i$ via $\Upsilon$.
\subsection{Notes for nonlinear $f$}\label{sec:nonlinearf} When $f$ is nonlinear,  $y \varphi_t(u,y)\,$ cannot be computed analytically---a key ingredient for our computational procedure. We approximate the nonlinear dynamics by linearizing $f$ around the mean of the initial distribution. Assuming that the initial state of the system is normally distributed as described by equation \eqref{eqn:linearGaussian}, linearizing around the mean of the initial state gives 
        \begin{align}\label{eqn:linearizedsystem}
              \x' \approx \mathbf{F} \cdot (\x - \overline{\x}_0) +f(\overline{\x}_0, 0) \,,
              \end{align}
      where $\mathbf{F}$ represents the Jacobian of $f$ at $t=0$ and $\x=\overline{\x}_0$; this reduces the nonlinear dynamical system to a form that is similar to equation \eqref{eqn:linearGaussian}. Thus, we can now use equations \eqref{eqn:gpx}, \eqref{eqn:gpvarphi}, and \eqref{eqn:compute_varphi} to approximate  $y \varphi_t(u,y)\,$ for nonlinear $f$.

\section{Machine-learned inverse maps}\label{sec:ml_inverse_maps}
\begin{figure}
    \centering
    \includegraphics[width=\textwidth]{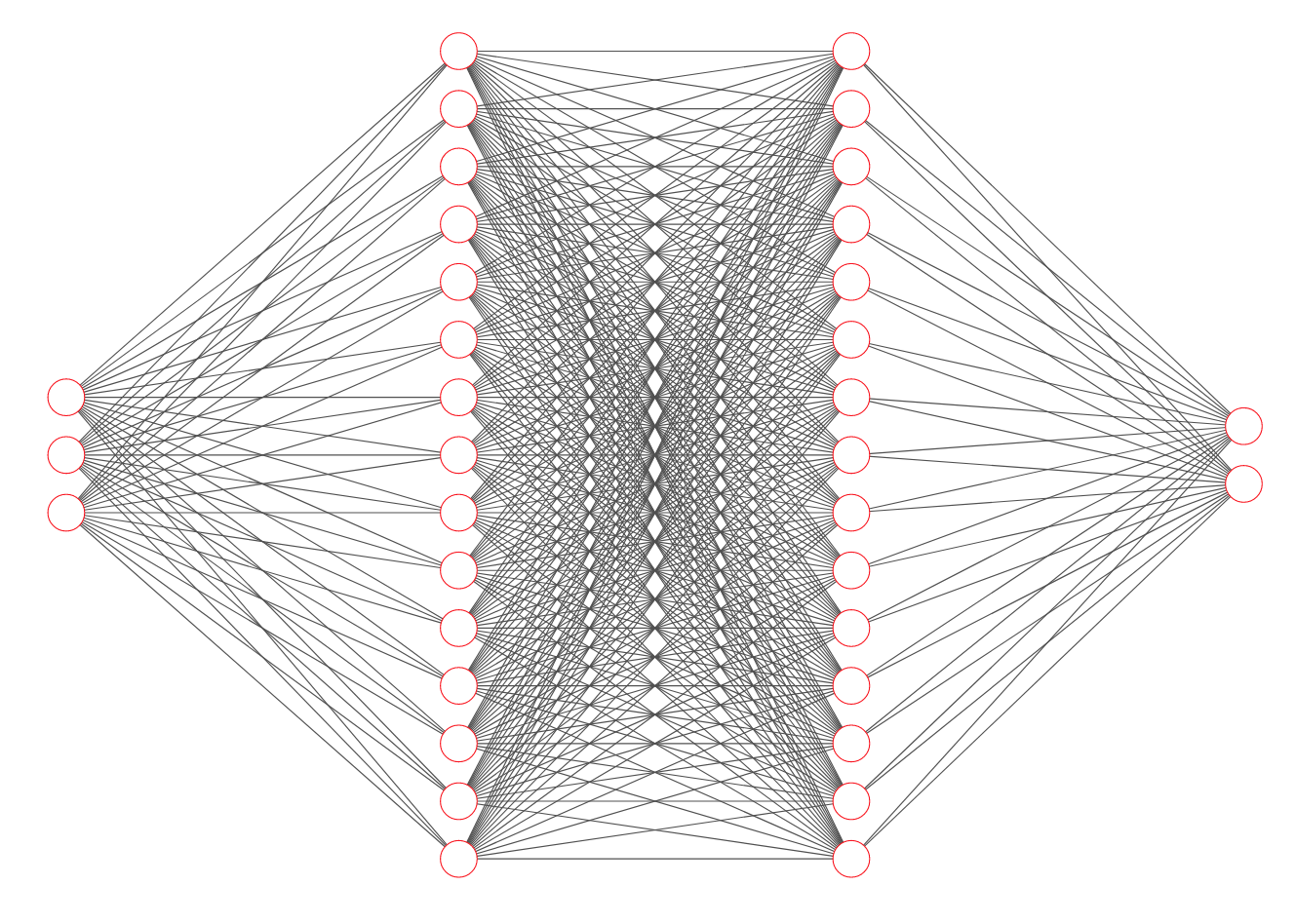}
    \caption{Schematic of our neural network architecture. Note that the number of hidden layer units are not representative since this study utilizes 256 such units. }
    \label{fig:NN_schematic}
\end{figure}

 In \cite{Rao_2020} we formulated the problem of determining preimages \eqref{eqn:preimages} as a Bayesian inverse problem. However, solving multiple Bayesian inverse problems can be expensive. Hence we approximated our IBD by using the solutions of a small number of Bayesian inverse problems. In this section we build a simple data-driven surrogate for approximating the preimages $X_i$ described in equation \eqref{eqn:preimages}. Using the surrogate, we can approximate the preimages of several $\y_i$'s obtained by sampling from $y\varphi_t(u,y)$. The surrogate developed here approximates the inverse of the map defined in equation \eqref{eqn:forwardmapG}. To that end, we wish to approximate the map 
\begin{align}
    \mathcal{G}^{-1} : \mathbb{R}^{2} \rightarrow \mathbb{R}^d,
\end{align}
where the input space corresponds to $(u+\varepsilon_i, y)\mid_{t_i}$ and the output lives in the domain of the state space ($\Omega$ here). This is equivalent to augmenting $t_i$ as an additional input variable and building a surrogate that maps from $\mathbb{R}^{3} \rightarrow \mathbb{R}^{d}$.
We utilize a fully connected deep neural network to approximate this map. A one-layered neural network can be expressed as
\begin{align}
    \label{DNN}
    \xi_{j}= F\left(\sum_{\ell=1}^{L} c_{m}^{\ell} x_{\ell}+\epsilon_{m}\right),
\end{align}
where $F$ is a differentiable activation function that imparts nonlinearity to this transformation; $L$ is the input dimension of an incoming signal; $M$ is the number of hidden-layer neurons (in machine learning terminology);  $c_m^{\ell} \in \mathbb{R}^{M \times L}$ are the weights of this map; $\epsilon_m \in \mathbb{R}^M$  are the biases; and $\xi_j \in \mathbb{R}^{J}$ is the nonlinear output of this map, which may be matched to targets available from data or ``fed-forward'' into future maps. Note that $\xi_j$ is the postactivation value of each neuron in a hidden layer of $J$ neurons. In practice, multiple compositions of this map may be used to obtain nonlinear function approximators,  called deep neural networks, that are very expressive. For   nonlinear activation, we utilize
\begin{align}
    F(\xi) = \text{max}(\xi,0),
\end{align}
for all its activation functions. In addition, we concatenate three such maps as shown in equation \ref{DNN} to ultimately obtain an approximation for $\mathcal{G}^{-1}$. Two such submaps have $J$ fixed at 256, and a final transformation utilizes $J=3$. We  note that the function $F$ for the final transformation is the identity, as is common in machine learning algorithms. A schematic of this network architecture is shown in Figure \ref{fig:NN_schematic}. The trainable parameters ($c_m^{\ell}$ and $\epsilon_m$ for each transformation) are optimized with the use of backpropagation \cite{rumelhart1986learning}, an adjoint calculation technique that obtains gradients of the loss function with respect to these parameters. A stochastic gradient optimization technique, ADAM, is used to update these parameters \cite{kingma2014adam} with a learning rate of 0.001. Our loss function is given by the $L_2$-distance between the prediction of the network and the targets (i.e., the mean-squared error). Our network also incorporates a regularization strategy, called dropout \cite{srivastava2014dropout}, that randomly switches off certain units $\xi_j$ (here we utilize a dropout probability of 0.1) in the forward propagation of the map (i.e., from $d \rightarrow 2$). Through this approach, memorization of data is avoided, while  allowing for effective exploration of a complex nonconvex loss surface.


\begin{figure}
    \centering
    \includegraphics[width=0.5\textwidth]{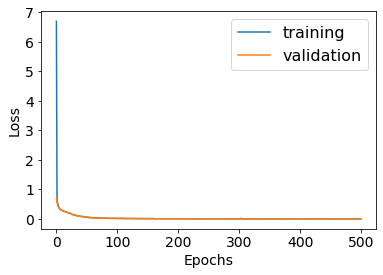}
    \caption{Convergence of training for our network. Note how both training and validation losses diminish in magnitude concurrently.}
    \label{losses}
\end{figure}

\begin{figure}
    \centering
    \mbox{
    \includegraphics[width=0.45\textwidth]{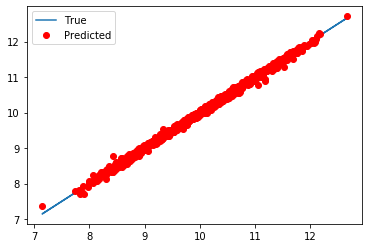}
    }
    \mbox{
    \includegraphics[width=0.45\textwidth]{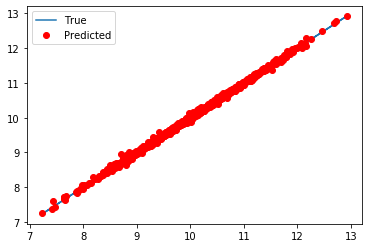}
    }
    \caption{Scatter plots between truth and predicted quantites of the inverse map with dimension 1 (left) and dimension 2 (right). These results are from unseen data.}
    \label{scatter}
\end{figure}

Our map is trained for 500 epochs with a batch size of 256; in other words,  a weight update is performed after a loss is computed for 256 samples. Each epoch is completed when the losses from the entire data set are used for gradient update. During the network training, we set aside a random subset of the data for  validation. Losses calculated from this data set are used only to monitor the learning of the framework for unseen data. These are plotted in Figure \ref{losses}, where one can  see that both training and validation losses are reduced to an equal magnitude. Figure \ref{scatter} also shows scatter plots for this validation data set where a good agreement between the true and predicted quantities can be seen. We may now use this map for approximating the IBD.

\subsection{Using the machine-learned inverse map to construct IBD}\label{sec:ml_ibd} The following procedure is used to construct the IBD.
\begin{enumerate}
    \item Obtain different realizations of the initial conditions of the dynamical system by sampling from the initial PDF $p$.
    \item Use $\mathcal{G}$ to obtain the forward maps of these realizations.
    \item Use the forward maps and the corresponding random realizations of the initial conditions to train the inverse map $\mathcal{G}^{-1}$.
    \item We now apply this trained inverse map on samples generated from $y \varphi_t(u,y)$ to obtain the approximate preimages of samples $\y_i$.
    \item Use a Gaussian approximation of these inverse maps is used as an IBD. Assume that this Gaussian approximation has PDF $p^{\rm IBD}$.
    \item Sample from the IBD, and use importance sampling to estimate the probabilities. 
\end{enumerate}
\subsection{Using IBD to estimate rare event probability} \label{sec:ml_is}
We can now estimate $P_T(u)$ using the IBD as follows:
\begin{align}\label{eqn:is_estimate}
        P_T^{\rm IS}(u)(\widehat{\mathbf{x}}_0^1, \ldots, \widehat{\mathbf{x}}_0^M) = \frac{1}{M} \sum_{i=1}^M\,\mathbb{I}(\widehat{\mathbf{x}}_0^i)\psi(\widehat{\mathbf{x}}_0^i)\,,
\end{align}
where $\widehat{\mathbf{x}}_0^1, \ldots, \widehat{\mathbf{x}}_0^M$ are sampled from the biasing distribution $p^{\rm IBD}$ and $\mathbb{I}(\widehat{\mathbf{x}}_0^i)$ represents the indicator function given by
\begin{align}
        \mathbb{I}(\widehat{\mathbf{x}}_0^i) =
\begin{cases}
1\,, \displaystyle \qquad \sup_{0 \leq t \leq T} \mathbf{c}^{\top}\mathbf{x}(t, \widehat{\mathbf{x}}_0^i) \geq u\,, ~~t\in \lbrack 0, T \rbrack\,,\\
0\,, \displaystyle \qquad \sup_{0 \leq t \leq T} \mathbf{c}^{\top}\mathbf{x}(t, \widehat{\mathbf{x}}_0^i) < u\,, ~~t\in \lbrack 0, T \rbrack\,.
\end{cases}
\end{align}
Also, $\psi(\widehat{\mathbf{x}}_0^i)$ represents the importance weights. The importance weight for an arbitrary $\widehat{\mathbf{x}}_0^i$ is given by
\begin{align}
     \displaystyle   \psi(\widehat{\mathbf{x}}_0^i) = \frac{p(\widehat{\x}_0^i)}{p^{\rm IBD}(\widehat{\x}_0^i)}\,.
\end{align}
\section{Numerical experiments}\label{sec:numerical_experiments}
\begin{figure}[]
  \centering
  \begin{minipage}[t]{0.48\textwidth}
    \includegraphics[width=\textwidth]{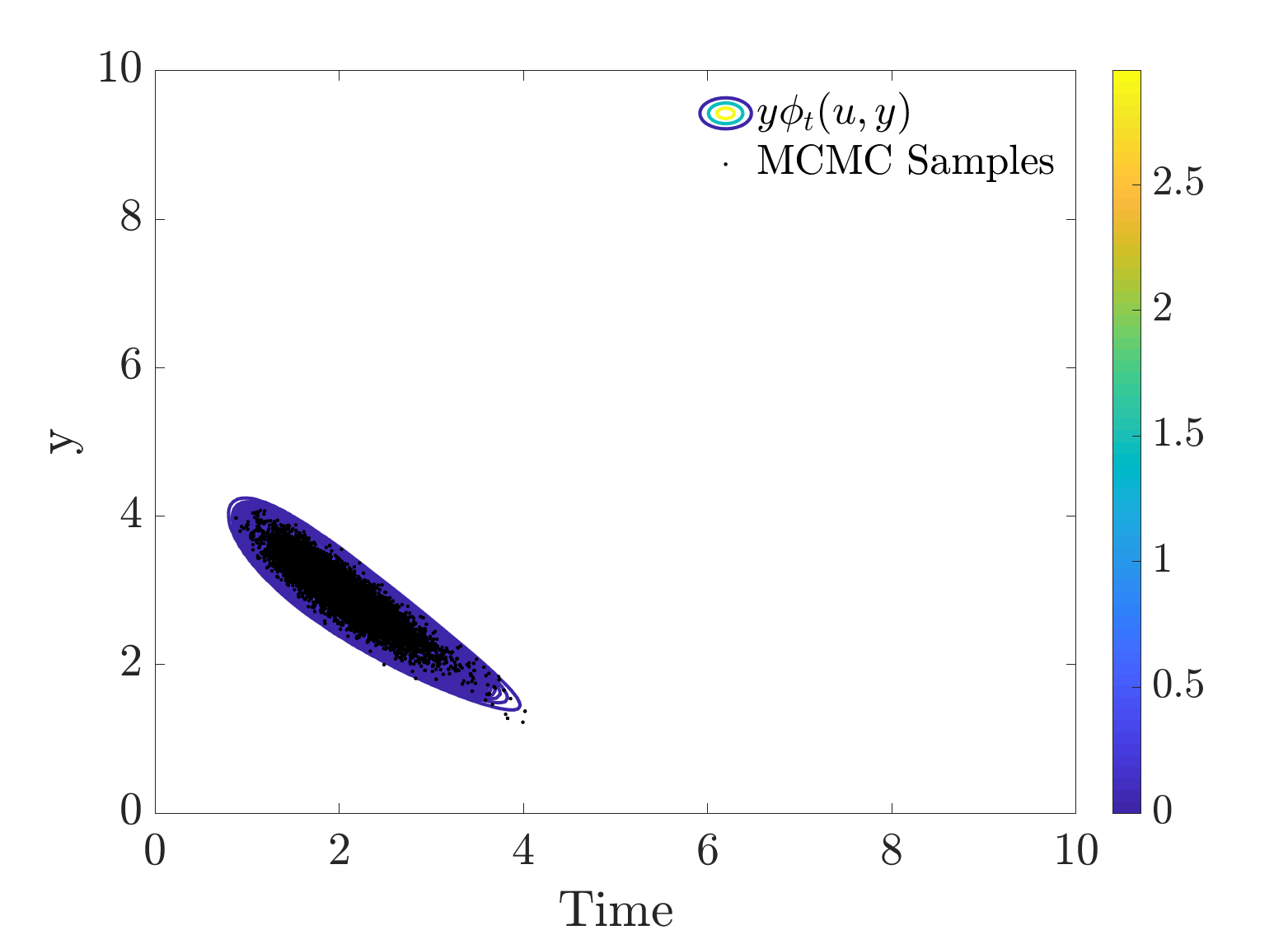}
    \caption{Contours of $y\varphi_t(u,y)$ and samples drawn from it using DRAM MCMC.}
    \label{fig:contoursofphi_U_20}
  \end{minipage}
  \hfill
  \begin{minipage}[t]{0.48\textwidth}
    \includegraphics[width=\textwidth]{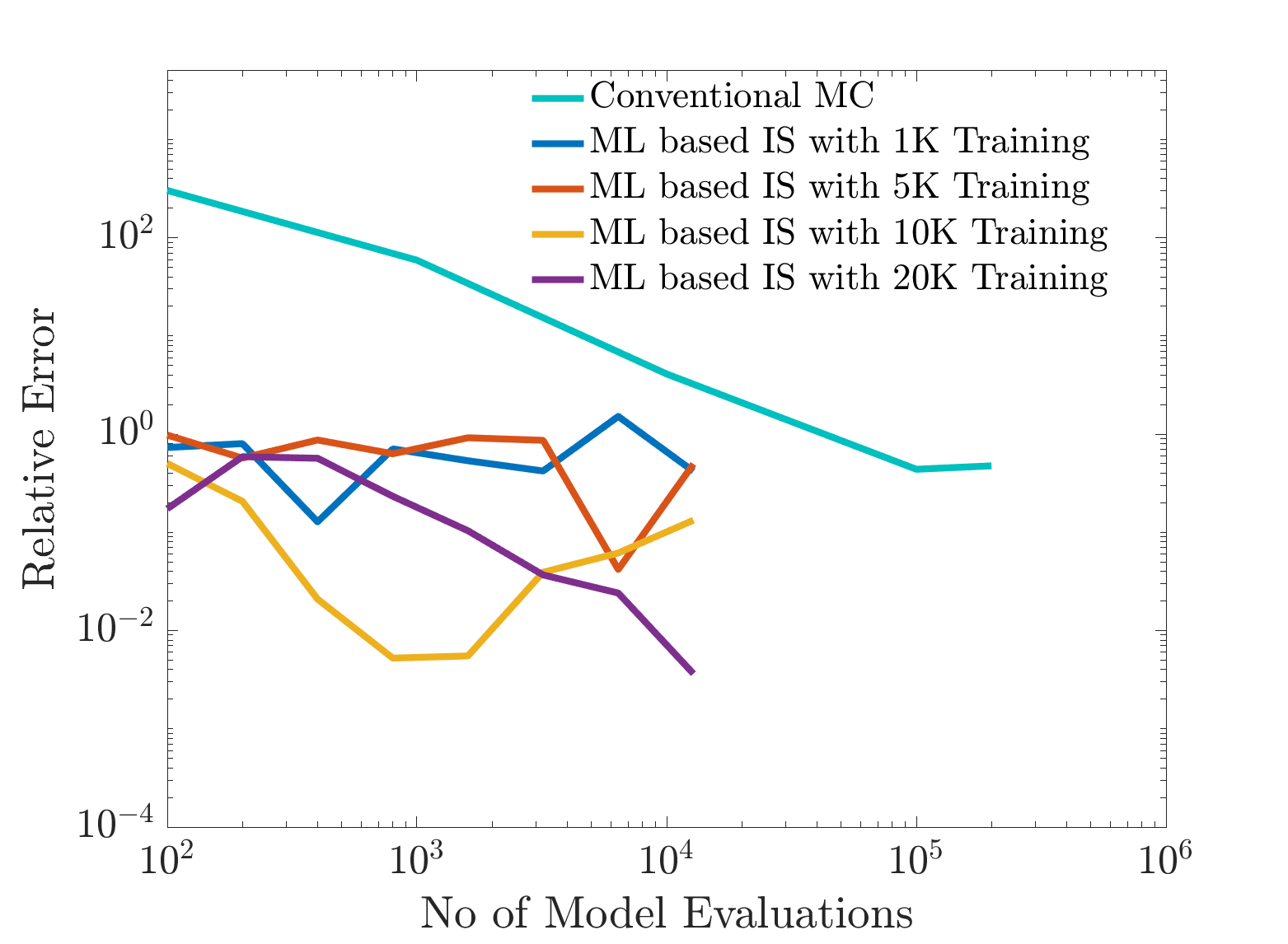}
    \caption{Comparison between conventional MCS and ML-based IS. We observe even with a small amount of training data, we obtain fairly accurate estimates; and as we increase the training data, the accuracy improves dramatically.}
    \label{fig:ML_vs_MCS}
  \end{minipage}
\end{figure}

We demonstrate the application ofthe procedure described in \Cref{sec:mcmc-based-sampling} and \Cref{sec:ml_inverse_maps} for nonlinear dynamical systems excited by a Gaussian distribution. We use the Lotka-Volterra equations as a test problem.
These equations,  also known as the predator-prey equations, are a pair of first-order nonlinear differential equations and are used to describe the dynamics of biological systems in which two species interact, one as a predator and the other as a prey. The populations change through time according to the following pair of equations,
\begin{align}\label{eqn:nonlinear_example}
\displaystyle \frac{dx_1}{dt} = \alpha x_1 - \beta x_1x_2 \,, \\
\nonumber \displaystyle \frac{dx_2}{dt} = \delta x_1x_2 - \gamma x_2\,,
\end{align}
where $x_1$ is the number of prey, $x_2$ is the number of predators, and $\displaystyle \frac{dx_1}{dt}$ and $\displaystyle \frac{dx_2}{dt}$ represent the instantaneous growth rates of the two populations. We assume that the initial state of the system at time $t=0$ is a random variable that is normally distributed:
$$\x(0) \sim \mathcal{N}\left(\begin{bmatrix}10 \\ 10 \end{bmatrix}, 0.8\times I_2\right),$$ and we are interested in estimating the probability of the event $P(\mathbf{c}^{\top}\x \geq u)$, where $\mathbf{c} = \begin{bmatrix} 0 \\ 1 \end{bmatrix}$, $t \in [0,10]$, and $u = 17$. The first step of our solution procedure involves sampling from $y \varphi_t(u,y)$ to generate observations $\y_i$. We  linearize the dynamical system about the mean of the distribution of $\x_0$ (equation \eqref{eqn:linearizedsystem}) and express $\varphi_t(u,y)$ as a function of $t$ and $y$ as described by equation \eqref{eqn:gpvarphi}. We  compute $y \varphi_t(u,y)$   as shown in equation \eqref{eqn:compute_varphi}. We use the delayed rejection adaptive Metropolis (DRAM) Markov chain Monte Carlo (MCMC) method to generate samples from $y \varphi_t(u,y)$. (For more details about DRAM,  see \cite{Haario_2006}.) To minimize the effect of the initial guess on the posterior inference, we use a burn-in of 1,000 samples. Figure \ref{fig:contoursofphi_U_20} shows the contours of $y \varphi_t(u,y)$ and samples drawn from it by using DRAM MCMC. 
In \cite{Rao_2020} we then
solved the Bayesian inverse problem by using both MCMC and Laplace approximation at MAP to construct a distribution that approximately maps to likelihood constructed around $\y_i$. 
Here, we replace the solution to the Bayesian inverse problem with a machine-learned inverse map described in \Cref{sec:ml_inverse_maps}. Multiple samples generated from $y\varphi_t(u,y)$ can be used to construct the IBD, as described in \S \ref{sec:ml_ibd}, and the IBD can be used to estimate $P_T(u)$, as explained in \S \ref{sec:ml_is}.    
Figure \ref{fig:ML_vs_MCS} compares the results between the conventional MCS and Machine Learning based Importance Sampling (ML-based IS) methods. Note that we use an MCS estimate with 10 Million samples as a proxy for the true probabilities. The ``true" probability is $3.28\times 10^{-5}$. ML-based IS gives fairly good estimate even with small number of model evaluations. When the training dataset size is large enough, the improvements are dramatic. Notice that for a true probability of the order of $10^{-5}$ we obtain an estimate that has a relative error of less than 1\%. Notice that our method gives the same (or better) accuracy as the MCS with hundred times lesser computational cost. The convergence with just 5000 training samples is acceptable and these results improve dramatically for 10000 and 20000 training samples. We believe the results could be even better when we use a Gaussian mixture to represent the IBD instead of a simple Gaussian approximation. 
\subsection{Computational Cost}In Figure \ref{fig:ML_vs_MCS}, we havent included the costs associated with generating training data, training costs, and cost for approximating the inverse map as these costs are almost negligible when compared to the overall costs. Note that generating 20000 samples is approximately equivalent to 400 model evaluations (this is because a single model evaluation can be used to generate the slope and state at 50 different times and each of them can be used as a training sample). The training of the ML framework, for this problem, required very little compute time. Each training was executed on an 8th-generation Intel Core-I7 machine with Python 3.6.8  and Tensorflow 1.14 and took less than 180 seconds for training 20000 samples (this is less than 50 model evaluations). Inference (for 20000 prediction points) costs were less than 2 seconds, on average. 

\section{Conclusions and future work}\label{sec:conc}
In this work we developed a ML-based IS to estimate rare event probabilities and we demostrated the algorithm on the prey-predator system. The method developed here builds on the approach in \cite{Rao_2020} and replaces the expensive Bayesian inference with a Machine learning based surrogate. This approach yields fairly accurate estimate of the probabilities and for a given accuracy requires atleast three orders of magnitude lesser computational effort than the traditional MCS. In future, we aim to test this algorithm for larger problems and also use an active learning based approach to pick the training samples. Scaling this algorithm to high dimensions (say $\mathcal{O}(1000)$) could be challenging and to address it, we will use state-of-the-art techniques developed by machine learning and deep learning community in the future. 
\bibliographystyle{splncs04}
\bibliography{References.bib}
%




\end{document}